\documentclass[10pt]{article}
\usepackage{graphicx}
\usepackage{lineno}

\def\Title#1{\begin{center} {\Large #1 } \end{center}}
\def\Author#1{\begin{center}{ \sc #1} \end{center}}
\def\Address#1{\begin{center}{ \it #1} \end{center}}

\newcommand\pubblock{\rightline{\begin{tabular}{l} Proceedings of the Fifth Annual LHCP\\ \pubnumber\\
         \pubdate  \end{tabular}}}

\newenvironment{Abstract}{\begin{quotation} \begin{center} 
             \large ABSTRACT \end{center}\bigskip 
      \begin{center}\begin{large}}{\end{large}\end{center} \end{quotation}}

\newenvironment{Presented}{\begin{quotation} \begin{center} 
             PRESENTED AT\end{center}\bigskip 
      \begin{center}\begin{large}}{\end{large}\end{center} \end{quotation}}

\def\Acknowledgements{\bigskip  \bigskip \begin{center} \begin{large}
             \bf ACKNOWLEDGEMENTS \end{large}\end{center}}
\def\beq{\begin{equation}}
\def\eeq#1{\label{#1}\end{equation}}
\def\eeqn{\end{equation}}

\def\beqa{\begin{eqnarray}}
\def\eeqa#1{\label{#1}\end{eqnarray}}
\def\eeqan{\end{eqnarray}}

\let\bar=\overbar

\def\Dslash{\not{\hbox{\kern-4pt $D$}}}
\def\dslash{\not{\hbox{\kern-2pt $\del$}}}

\def\msb{{\bar{\ssstyle M \kern -1pt S}}}

\usepackage{color}

\newcommand\pubnumber{ ATL-PHYS-PROC-2017-104 }

\newcommand{\lb}  {\mbox{$\Lambda_b^0$}}
\newcommand{\prob}{\mbox{$\cal P$}}
\newcommand{\bran}{\mbox{$\cal B$}}
\newcommand{\pt}{\mbox{$p_{\rm T}$}}

\newcommand\pubdate{\today}

\def\affiliation{
On behalf of the ATLAS Collaboration, \\
Lomonosov Moscow State University Skobeltsyn Institute of Nuclear Physics (SINP MSU), \\
Moscow 119991, Russian Federation}

\begin{document}

% large size for the first page
\large
\begin{titlepage}
\pubblock

%% Change the title, name, abstract
%% Title 
\vfill
\Title{  ATLAS results on hadron spectroscopy, including exotic states  }
\vfill

%  if you need to add the support use this, fill the \support definition above. 
%   \Author{ FIRSTNAME LASTNAME \support }
\Author{ Leonid Gladilin  }
\Address{\affiliation}
\vfill
\begin{Abstract}

  Recent results of the ATLAS experiment at LHC on hadron spectroscopy,
  including exotic states,
  are presented.
  Comparison of the results
  with various theoretical predictions is discussed.

\end{Abstract}
\vfill

% DO NOT CHANGE 
\begin{Presented}
The Fifth Annual Conference\\
 on Large Hadron Collider Physics \\
Shanghai Jiao Tong University, Shanghai, China\\ 
May 15-20, 2017
\end{Presented}
\vfill
\end{titlepage}
\def\thefootnote{\fnsymbol{footnote}}
\setcounter{footnote}{0}
%

% normal size for the rest
\normalsize 

%% Your paper should be entered below. 
%\linenumbers
%

\section{Introduction}

The ATLAS detector~\cite{atldet} at the Large Hadron Collider (LHC)
consists of several subsystems including the inner
detector (ID), the electromagnetic and hadronic calorimeters,
and the muon spectrometer (MS). Muon reconstruction at ATLAS
makes use of both the ID and the MS,
and covers the pseudorapidity range $|\eta|<2.5$
A three-level trigger system allows ATLAS
to effectively select events
containing single muons with large transverse momentum $\pt$ (above $\sim 20\,$GeV),
events with two muons with
the minimal muon $\pt$ thresholds of $4-6\,$GeV,
and events with three muons
with the muon $\pt$ threshold of $4\,$GeV.
During 2011-2012 data taking ATLAS has accumulated data samples
of $pp$ collisions
corresponding to luminosities of $\sim 5\,$fb$^{-1}$ at $\sqrt{s}=7\,$TeV
and $\sim 20\,$fb$^{-1}$ at $\sqrt{s}=8\,$TeV.

In this proceeding,
recent results of the ATLAS experiment at LHC on hadron spectroscopy,
including exotic states, are presented.
The $\lb \rightarrow \psi(2S) \Lambda^0$ decay
is observed
and
the branching ratio
of the
$\lb \rightarrow \psi(2S) \Lambda^0$ and
$\lb \rightarrow J/\psi \Lambda^0$ decays is measured~\cite{atllbpsi2s}.
The decays
$B_c^+\to J/\psi D_s^+$ and
$B_c^+\to J/\psi D_s^{*+}$ are studied
and their branching fractions are measured relative
to that of the $B_c^+\to J/\psi \pi^+$ decay~\cite{atlbcds}.
The production cross sections and properties of
the hidden-charm states $X(3872)$ and $\psi(2S)$ are measured
in their decays to $J/\psi \pi^+\pi^-$~\cite{atlx3872}.
A  search  for  a  hidden-beauty  analogue  of  the
$X(3872)$, $X_b$,  is
conducted  by  reconstructing
$\Upsilon(1S)(\to \mu^+\mu^-) \pi^+\pi^-$ events~\cite{atlxb}.
Corrections for detector effects are done
with high-statistics Monte Carlo (MC) samples.
Uncertainties due to simulation of physics processes and detector,
MC statistic, luminosity measurement and assumptions
of the analysis procedures are included into systematic errors.
The measurements are compared to theoretical
predictions and to the measurements by other experiments.

\section{Observation of the $\lb \rightarrow \psi(2S) \Lambda^0$ decay}

In all events with $J/\psi$ or $\psi(2S)$
candidates, pairs of tracks from particles with opposite charge are
combined to form $\Lambda^0$ candidates.
Tracks of the selected charmonium and $\Lambda^0$
candidates are simultaneously refitted with the dimuon and dihadron masses
constrained to the world average masses of $J/\psi$ or $\psi(2S)$
and $\Lambda^0$~\cite{pdg2016}, respectively.
The combined momentum of the refitted $\Lambda^0$
track pair is required to point to the dimuon vertex.
To control $B^0$
reflections to the $\Lambda_b^0$ 
signal distributions, a $B^0$
decay topology fit is also attempted
for each track quadruplet successfully fitted to the $\Lambda_b^0$
topology, i.e.  the pion mass is assigned to both
hadron tracks and the dihadron mass is constrained to
the world average mass of $K^0_S$~\cite{pdg2016}.
To suppress the $B^0$ background
the requirement $\prob(\Lambda_b^0) > \prob(B^0)$
is applied, where
$\prob(\Lambda_b^0)$ and $\prob(B^0)$
are the $\chi^2$
probabilities of the quadruplet fits with
$\Lambda_b^0$ and $B^0$
topologies, respectively.

\begin{figure}[htb]
\centering
\includegraphics[height=2in]{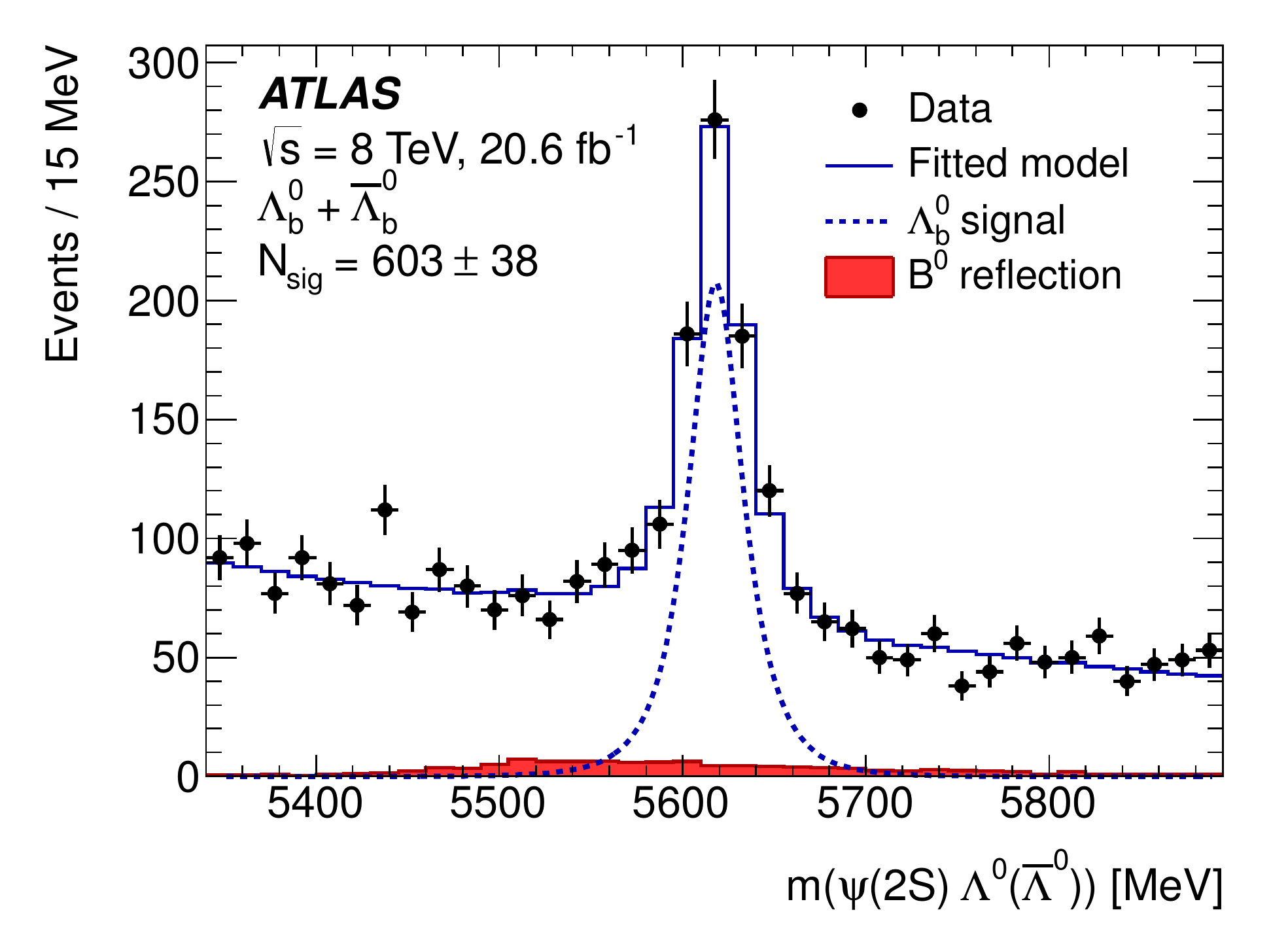}
\includegraphics[height=2in]{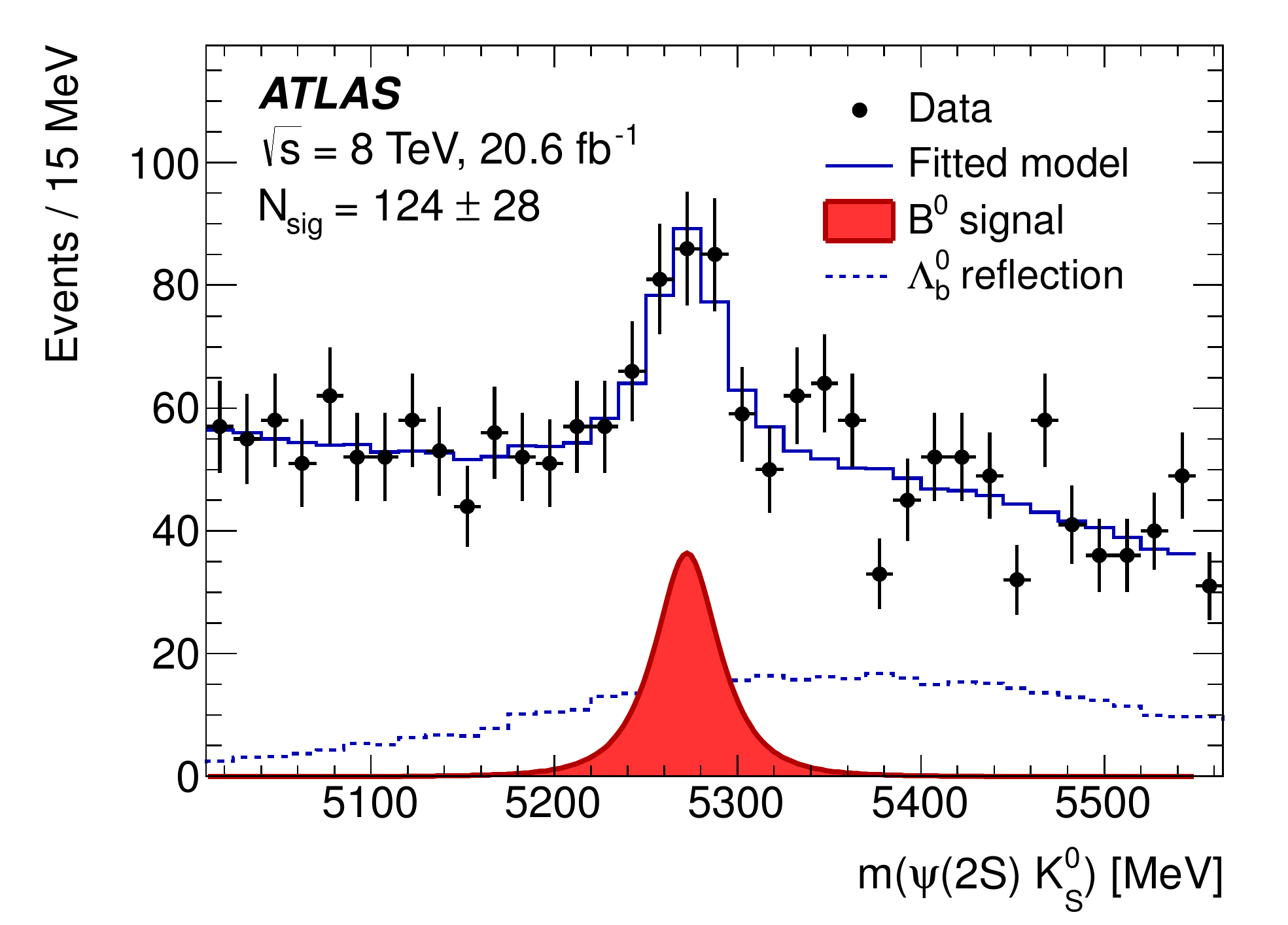}
\caption{The invariant mass distributions for the combined sample of the selected $\Lambda_b^0$ and ${\bar \Lambda}_b^0$ candidates obtained after their fits to the $\Lambda_b^0 \to \psi(2S)\Lambda^0$ (left plot) and $B^0 \to \psi(2S) K_S^0$ (right plot) topologies~\cite{atllbpsi2s}. The solid histograms represent fit results. The $\Lambda_b^0$ and $B^0$ signals and their mutual reflections are also shown.}
\label{fig:fig_psi2s}
\end{figure}

Figure~\ref{fig:fig_psi2s} shows
the invariant mass distributions for the combined sample
of the selected $\Lambda_b^0$ and ${\bar \Lambda}_b^0$ candidates
obtained after their fits to the $\Lambda_b^0 \to \psi(2S)\Lambda^0$
and $B^0 \to \psi(2S) K_S^0$ topologies.
Clear $\Lambda_b^0$ and $B^0$ signals are seen.
The $m(\psi(2S)\Lambda^0)$ and $m(\psi(2S)K^0_S)$
distributions are simultaneously fitted to sums of signal and two-component
background distributions. The signals are described by modified Gaussian
functions~\cite{modgauss}.
The non-resonant backgrounds
are described by independent exponential functions.
The mutual $B^0$ and $\Lambda^0_b$
reflections
are described by MC templates normalised to the numbers of
$B^0$ and $\Lambda^0_b$
hadrons obtained in the fit.
The fit yields $N(\Lambda_b^0 \to \psi(2S)\Lambda^0) = 603 \pm 38$.
A similar fit of the $m(J/\psi(2S)\Lambda^0)$ and $m(J/\psi(2S)K^0_S)$
distributions yields $N(\Lambda_b^0 \to J/\psi(2S)\Lambda^0) = 6940 \pm 130$.
Using these yields and correcting for detector effects and for
the branching fractions of the $J/\psi$ and $\psi(2S)$ decays
to two muons,
the branching ratio
of the
$\lb \rightarrow \psi(2S) \Lambda^0$ and
$\lb \rightarrow J/\psi \Lambda^0$ decays is measured to be
\begin{linenomath}
  $$\frac{\Gamma(\lb \rightarrow \psi(2S)\Lambda^0)}{\Gamma(\lb \rightarrow J/\psi\,\Lambda^0)}=0.501\pm 0.033\, ({\rm stat})\pm 0.016\, ({\rm syst})\pm 0.011\, ({\bran}),$$
\end{linenomath}
where
the third uncertainty originates from
the uncertainties of the charmonium
branching fractions.

The measured ratio
lies in the range 0.5--0.8 found for the branching ratios
of analogous $B$ meson decays~\cite{pdg2016}. 
The only available calculation for the branching ratio of the two $\lb$ decays
($0.8\pm 0.1$~\cite{gutsche,gutsche2}) exceeds the measured value.

\section{Study of the  $B_c^+\to J/\psi D_s^+$ and
$B_c^+\to J/\psi D_s^{*+}$ decays}

For the
$D_s^+\to \phi(K^+K^-)\pi^+$
reconstruction, tracks of particles with opposite charges are assigned kaon
mass hypotheses and combined in pairs to form $\phi$
candidates.   An additional track is assigned a pion
mass and combined with the $\phi$
candidate to form a
$D_s^+$
candidate.

\begin{figure}[htb]
\centering
\includegraphics[height=2in]{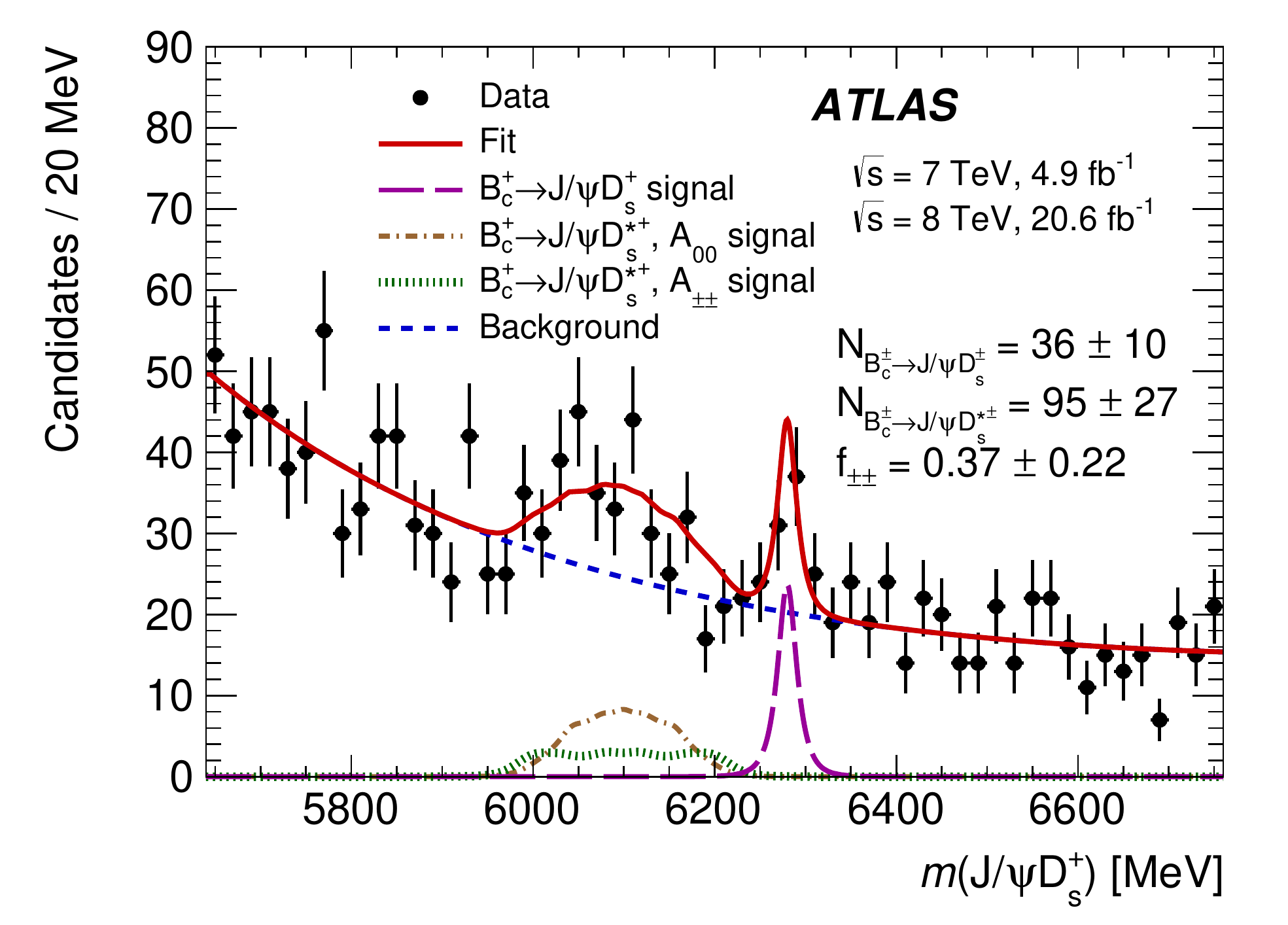}
\includegraphics[height=2in]{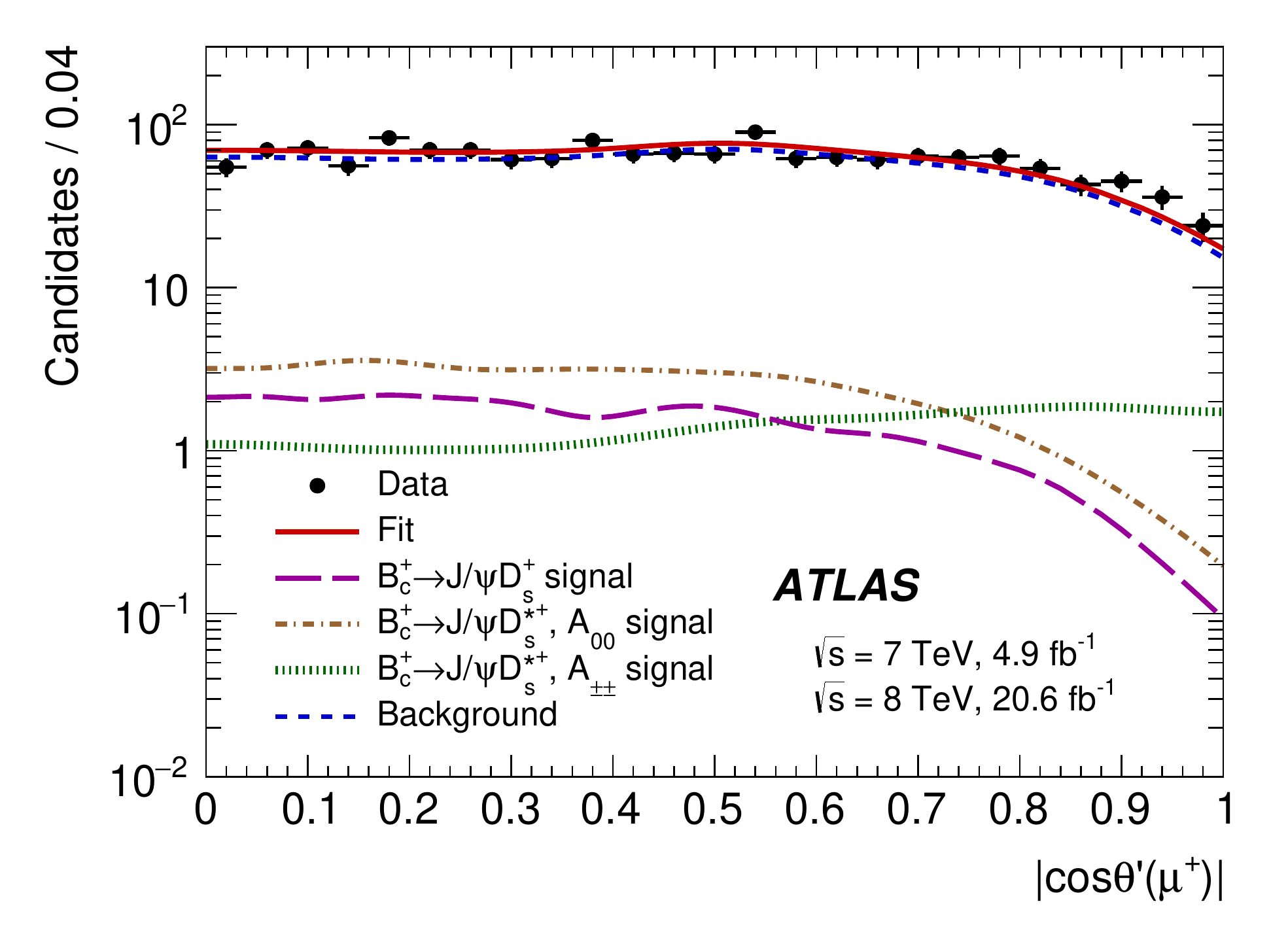}
\caption{The invariant mass distribution for the selected $J/\psi D_s^+$ candidates (left plot)
  and the $|\cos \theta^\prime(\mu^+)|$ distribution (right plot), where the helicity angle $\theta^\prime(\mu^+)$
  is the angle between the $\mu^+$ and $D_s^+$ candidate momenta in the rest frame of the muon pair
  from $J/\psi$ decay~\cite{atlbcds}.
  The red solid lines represent the projection of the likelihood fit to the model described in the text.
  The contributions of the $B_c^+\to J/\psi D_s^+$ decay are shown with the magenta long-dashed lines;
  the brown dash-dot and green dotted lines show the $B_c^+\to J/\psi D_s^{*+}$ $A_{00}$ and $A_{\pm\pm}$ component contributions, respectively;
  the blue dashed lines show the background model.}
\label{fig:fig_bcds}
\end{figure}

\begin{figure}[htb]
\centering
\includegraphics[height=2in]{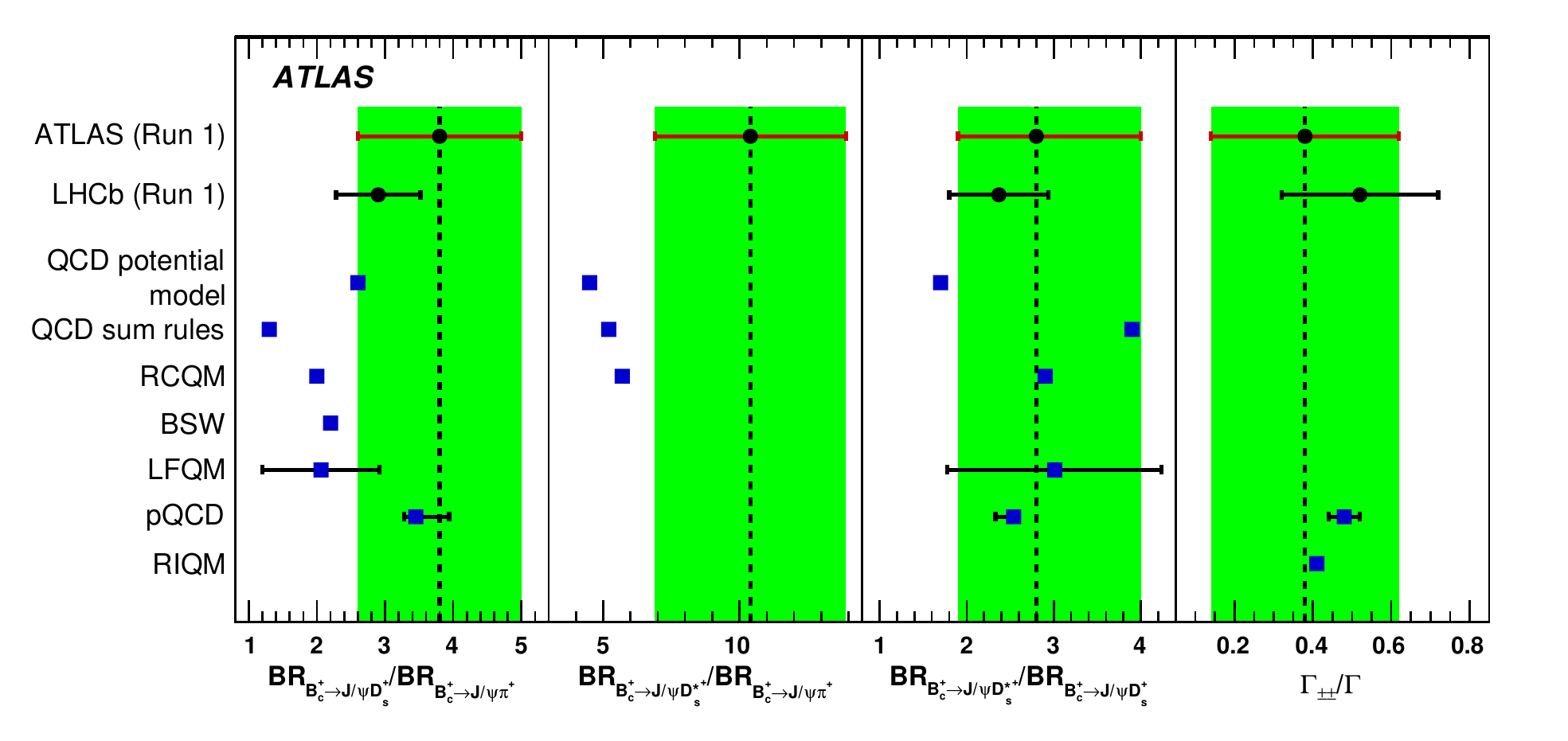}
\caption{Comparison of the results of the ATLAS measurement with those of LHCb
  and theoretical predictions based on a QCD relativistic potential model,
  QCD sum rules, relativistic constituent quark model (RCQM), BSW relativistic quark model, light-front quark model,
  perturbative QCD (pQCD), and relativistic independent quark model (RIQM)~\cite{atlbcds}.
  %  The uncertainties of the theoretical predictions are shown if they are explicitly quoted in the corresponding papers. Statistical and systematic uncertainties added in quadrature are quoted for the results of ATLAS and LHCb.
}
\label{fig:fig_bcds_res}
\end{figure}

The $B_c^+\to J/\psi D_s^+$
candidates are built by combining the five tracks of the
$J/\psi$ and $D_s^+$ candidates.
Figure~\ref{fig:fig_bcds} shows
the invariant mass distribution for the selected $J/\psi D_s^+$ candidates
and the $|\cos \theta^\prime(\mu^+)|$ distribution,
where the helicity angle $\theta^\prime(\mu^+)$
is the angle between the $\mu^+$ and $D_s^+$ candidate momenta
in the rest frame of the muon pair
from $J/\psi$ decay.
The peak near
the $B_c^+$
is attributed to the signal of
$B_c^+\to J/\psi D_s^+$
decay while a wider
structure between 5900 MeV and 6200 MeV corresponds to
$B_c^+\to J/\psi D_s^{*+}$
with subsequent
$D_s^{*+}\to D_s^+\gamma$ or
$D_s^{*+}\to D_s^+\pi^0$
decays where the neutral particle is not reconstructed.
A two-dimensional
extended unbinned maximum-likelihood fit of the
$m(J/\psi D_s^+)$ and
$|\cos\theta^\prime(\mu^+)|$
distributions is performed using
four two-dimensional probability density functions (PDFs) to describe the
$B_c^+\to J/\psi D_s^+$
signal, the
$A_{\pm\pm}$ and $A_{00}$ 
components of the
$B_c^+\to J/\psi D_s^{*+}$
signal, and the background.
The helicity amplitudes
$A_{++}$, $A_{--}$ and $A_{00}$
correspond to the helicities of
$J/\psi$ and
$D_s^{*+}$ mesons.
The mass distribution of the
$B_c^+\to J/\psi D_s^+$
signal is described by a modified Gaussian function.  For the
$B_c^+\to J/\psi D_s^{*+}$
signal components, the mass shape templates obtained from
the simulation with the kernel
estimation technique are used.
The transverse polarisation fraction is determined to
be  $\Gamma_{\pm\pm}(B_c^+\to J/\psi D_s^{*+})/\Gamma(B_c^+\to J/\psi D_s^{*+}) =
0.38\pm 0.23 \pm 0.07$, and the derived ratio of the
branching fractions of the two modes is
$\bran_{B_c^+\to J/\psi D_s^{*+}}/\bran_{B_c^+\to J/\psi D_s^+} =
2.8^{+1.2}_{-0.8} \pm 0.3$.
A sample of
$B_c^+\to J/\psi\pi^+$
decays
is used to derive the ratios of branching fractions
$\bran_{B_c^+\to J/\psi D_s^{+}}/\bran_{B_c^+\to J/\psi \pi^+} =
3.8 \pm 1.1 \pm 0.4 \pm 0.2$ and
$\bran_{B_c^+\to J/\psi D_s^{*+}}/\bran_{B_c^+\to J/\psi \pi^+} =
10.4 \pm 3.1 \pm 1.5 \pm 0.6$,
where the third error corresponds to the
uncertainty of the branching fraction of
$D_s^+\to \phi(K^+K^-)\pi^+$
decay.
Figure~\ref{fig:fig_bcds_res} compares
these  results with  those  of  the  LHCb  measurement~\cite{lhcbbcds}  and
to the expectations from
various theoretical calculations.
The measured ratios of the branching fraction are generally
described by perturbative QCD,
sum rules, and relativistic quark models.
There is an indication of underestimation of the decay rates
for the
$B_c^+\to J/\psi D_s^{*+}$
decays by some models, although the discrepancies do not exceed two standard
deviations when taking into account only the experimental uncertainty.
The measurement results agree
with those published by the LHCb experiment.

\section{Measurement of the hidden-charm states $X(3872)$ and $\psi(2S)$}

Two muon tracks are fitted to a common vertex.
The dimuon
invariant mass is then constrained to the
$J/\psi$ mass, and the four-track vertex fit of the two muon tracks and
pairs of non-muon tracks is performed to find
$J/\psi \pi^+\pi^-$
candidates.
Only $J/\psi \pi^+\pi^-$
combinations with rapidity
within the range
$|y|<0.75$ and
transverse momenta
within the range
$10 < \pt < 70\,$GeV
are considered.

\begin{figure}[htb]
\centering
\includegraphics[height=2in]{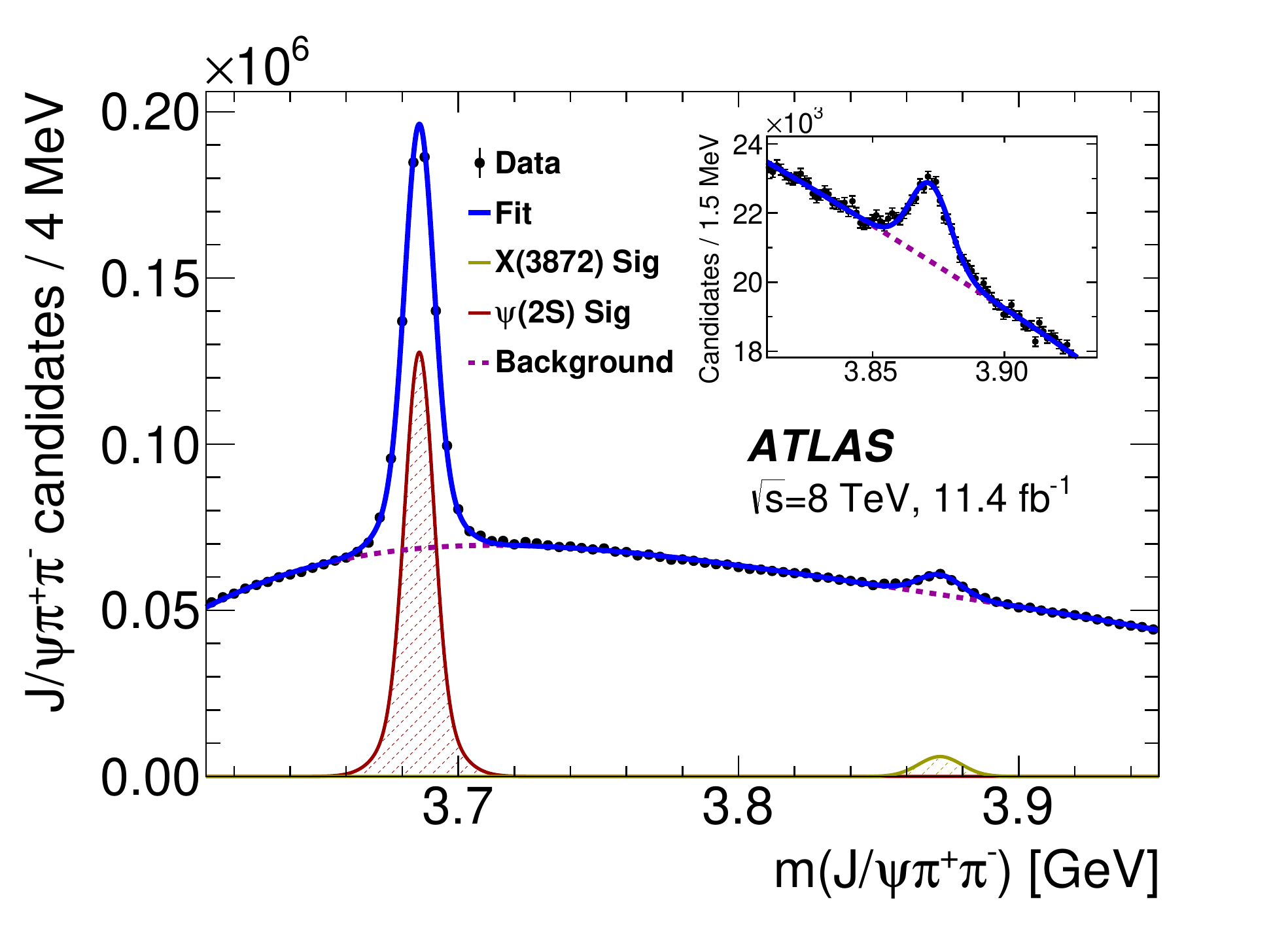}
\includegraphics[height=2in]{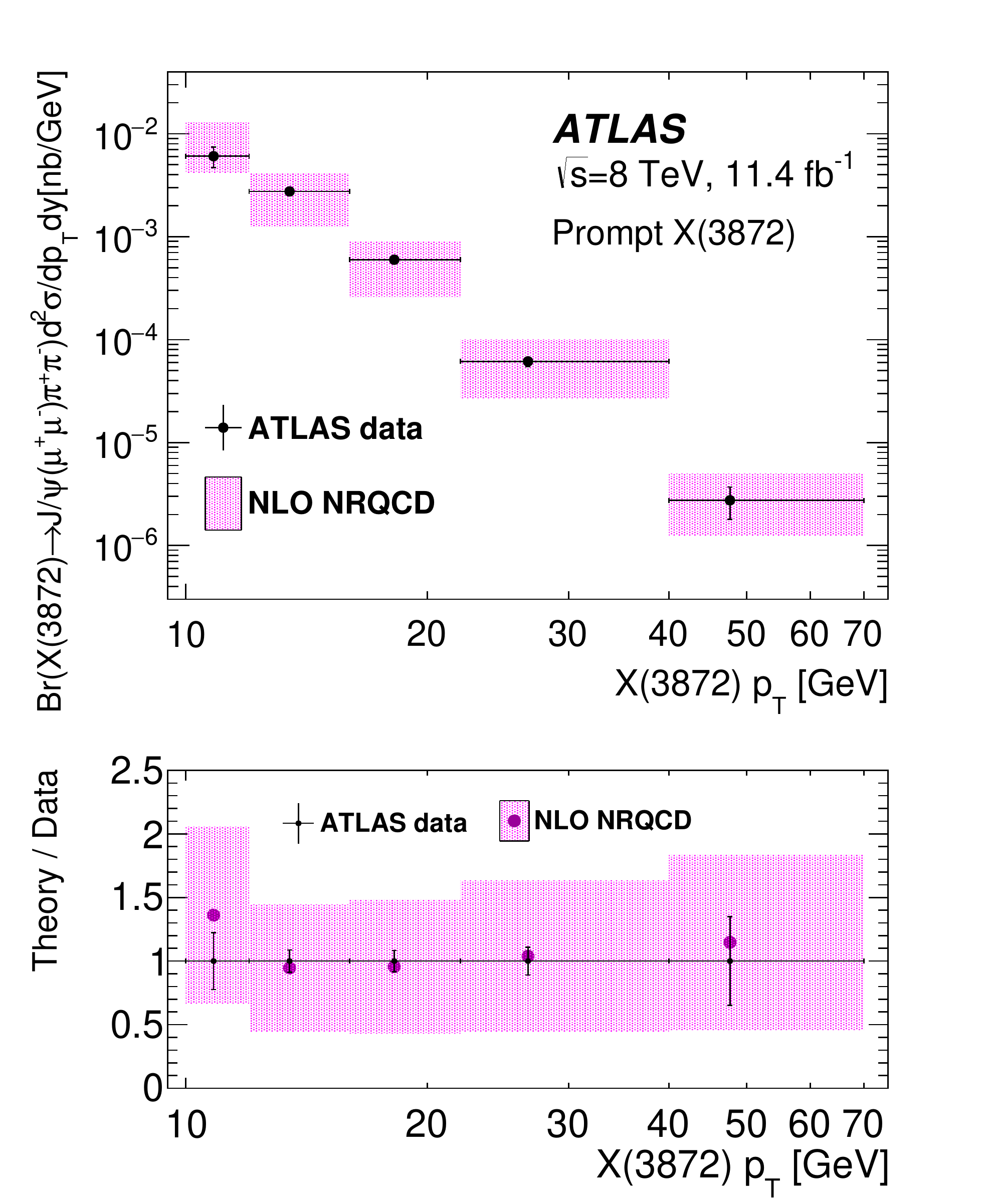}
\caption{(Left) Invariant mass of the selected $J/\psi \pi^+\pi^-$ candidates collected over the full $\pt$ range 10--70$\,$GeV and the rapidity range $|y|<0.75$~\cite{atlx3872}. The curve shows the results of the fit using double-Gaussian functions for the $\psi(2S)$ and $X(3872)$ peaks and a fourth-order polynomial for the background. The $X(3872)$ mass range is highlighted
in the inset.
(Right) Measured cross section times branching fractions as a function of $\pt$ for prompt $X(3872)$ production~\cite{atlx3872} compared to NLO NRQCD predictions~\cite{meng}.
%with the $X(3872)$ modelled as a mixture of $\chi_{c1}(2P)$ and a $D^0 {\bar D}^{*0}$ molecular state~\cite{meng}.
}
\label{fig:fig_x3872_sig}
\end{figure}

Figure~\ref{fig:fig_x3872_sig}(Left) shows
the invariant mass of the selected $J/\psi \pi^+\pi^-$ candidates.
The fitted function is the sum of a fourth-order polynomial background and two
double-Gaussian functions.  The double-Gaussian functions for
$\psi(2S)$ and $X(3872)$
contain about 470 k
and 30 k candidates, respectively.
To separate prompt production of the
$\psi(2S)$ and $X(3872)$
states from the non-prompt production
occurring via the decays of long-lived particles such as
b-hadrons,
the pseudo-proper lifetime
$\tau$ is used.
The pseudo-proper lifetime
is defined as
$\tau = L_{xy} m / c \pt$,
where
$L_{xy}$
is the transverse decay length,
$m$
is the invariant mass and
$\pt$
is the transverse momentum of
the $J/\psi \pi^+\pi^-$
candidate.

The measured differential cross section
(times the product of the relevant branching fractions)
for prompt
production of $X(3872)$
is shown in Figure~\ref{fig:fig_x3872_sig}(Right).
It is described within the theoretical uncertainty by the prediction
of the NRQCD model which, in this case,
considers $X(3872)$ to be a mixture of
$\chi_{c1}(2P)$
and a
$D^0{\bar D}^{*0}$
molecular state~\cite{meng}.
However, the prediction for the $X(3872)$ prompt production is
dominated by the
$\chi_{c1}(2P)$
component.

\begin{figure}[htb]
\centering
\includegraphics[height=2in]{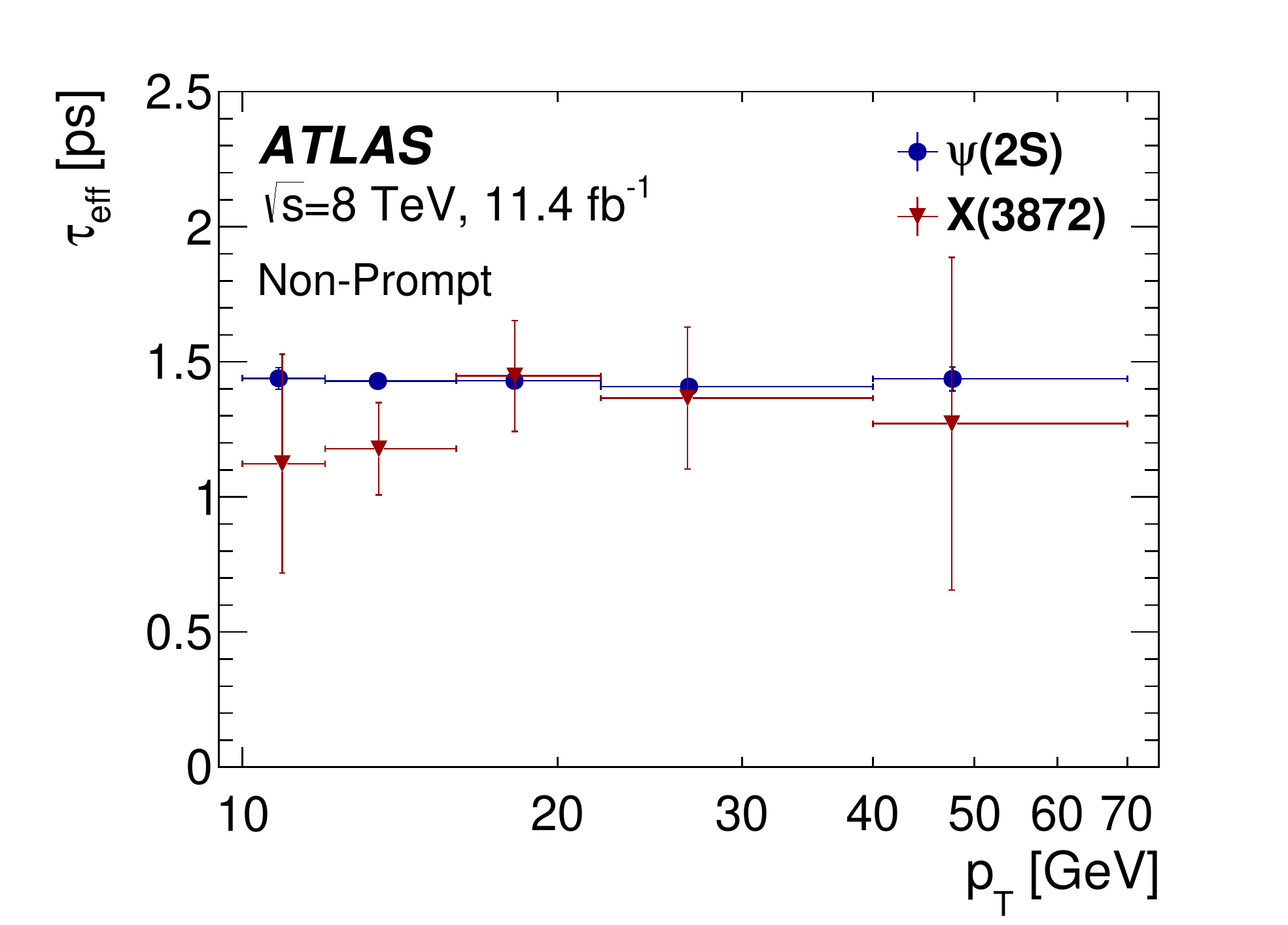}
\includegraphics[height=2in]{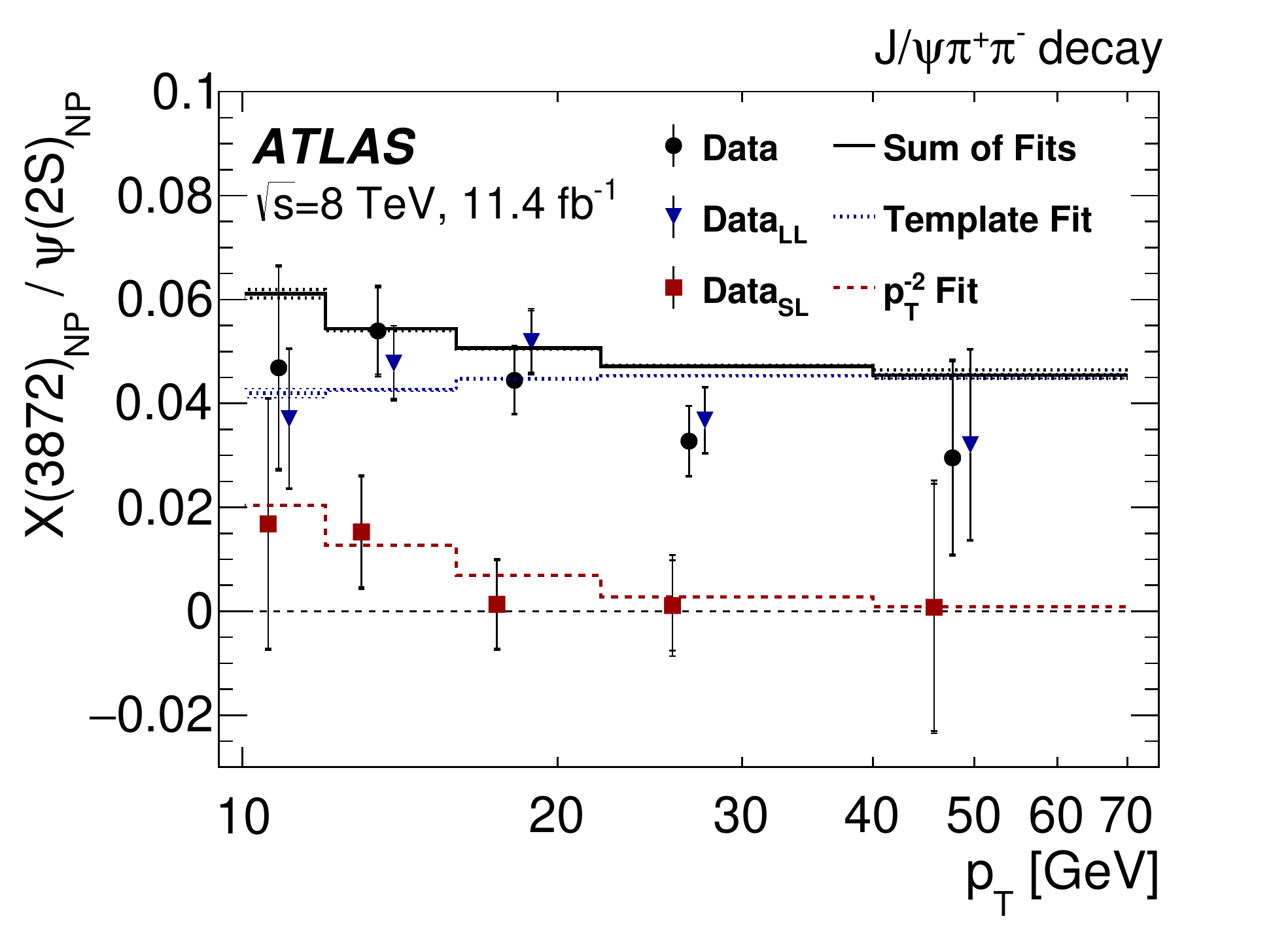}
\caption{(Left) Measured effective pseudo-proper lifetimes for non-prompt $X(3872)$ and $\psi(2S)$~\cite{atlx3872}.
  (Right) Ratio of cross sections times branching fractions, $X(3872) / \psi(2S)$, for the total non-prompt ratio (black circles), short-lived (red squares) and long-lived (blue triangles) components for the $X(3872)$, shown with respective fits described in the text~\cite{atlx3872}. The data points are slightly shifted horizontally for visibility.
}
\label{fig:fig_x3872_tau}
\end{figure}

Figure~\ref{fig:fig_x3872_tau}(Left) shows
the measured effective pseudo-proper lifetimes for non-prompt $X(3872)$ and $\psi(2S)$ in bins of $\pt$.
While for $\psi(2S)$
the fitted values of $\tau_{\rm eff}$
are measured to be around $1.45\,$ps
in all $\pt$ bins, the signal from $X(3872)$ at low
$\pt$
tends to have shorter lifetimes, possibly hinting at a
different production mechanism at low $\pt$
with the short-lived part
due to the contribution of
$B^\pm_c$ mesons.
To study this the non-prompt
production cross section of $X(3872)$ is split
into short-lived ($\tau_{\rm SL}=0.40\pm 0.05\,$ps) and long-lived
($\tau_{\rm LL}=1.45\pm 0.05\,$ps) components.
The  ratio  of  short-lived  non-prompt
$X(3872)$ to  non-prompt $\psi(2S)$,
shown  in  Figure~\ref{fig:fig_x3872_tau}(Right),
is fitted with a function
$a/p^2_T$~\cite{berezhnoy}.
The value of $a$,
and the measured non-prompt yields of
$X(3872)$ and $\psi(2S)$ states, are used to determine
the fraction of non-prompt
$X(3872)$ from short-lived sources, integrated over the
$\pt$ range ($\pt > 10\,$GeV)
covered in this measurement, giving:
\begin{linenomath}
$$\frac{\sigma(pp \to B_c + {\rm any})\bran(B_c \to X(3872)+{\rm any})}
{\sigma(pp \to X(3872)+{\rm any})} = (25\pm 13\,({\rm stat}) \pm 2\,({\rm sys}) \pm 5\,({\rm spin}))\%,
$$
\end{linenomath}
where the last uncertainty comes from varying the spin-alignment of
$X(3872)$. 

\begin{figure}[htb]
\centering
\includegraphics[height=2in]{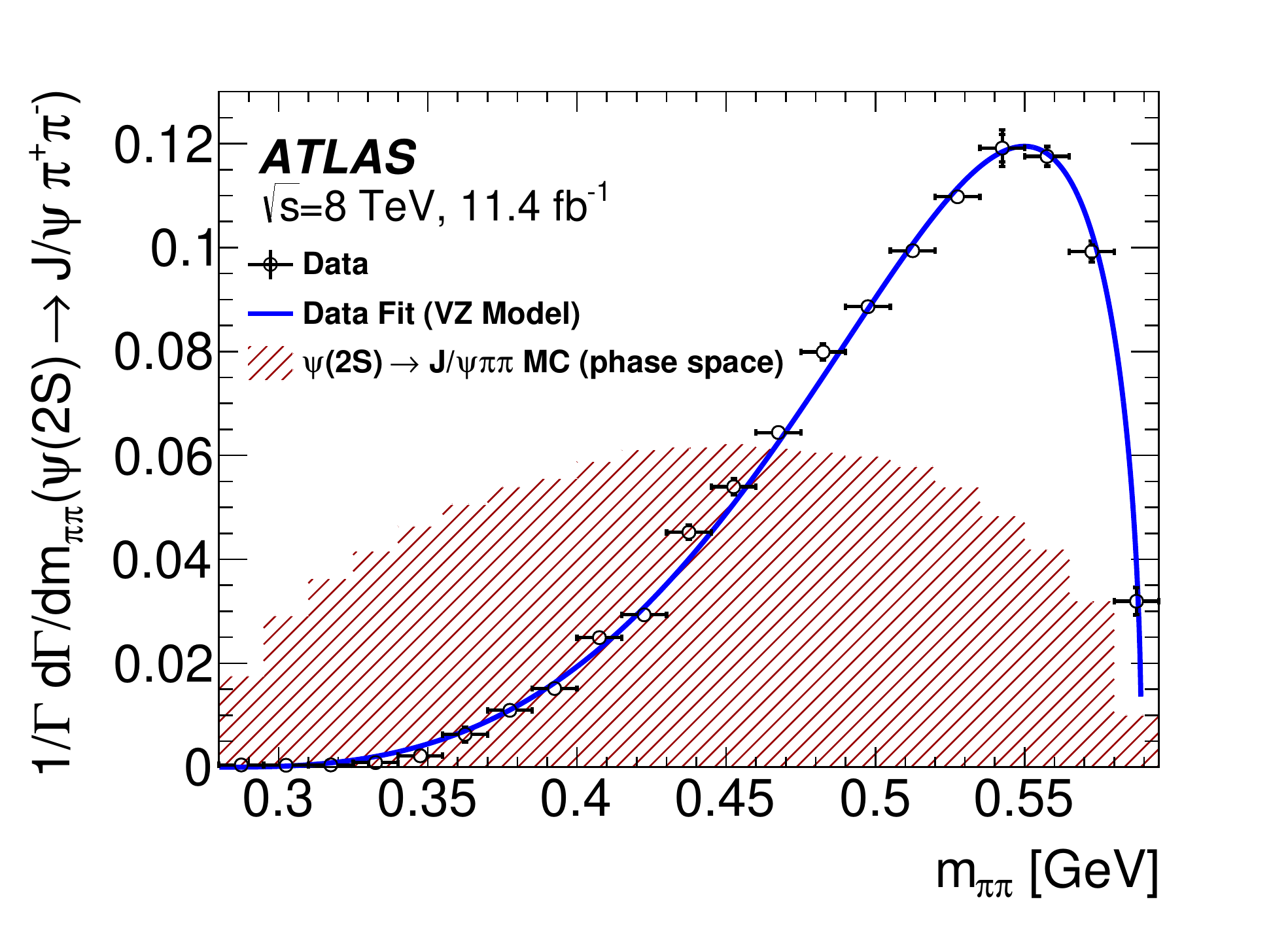}
\includegraphics[height=2in]{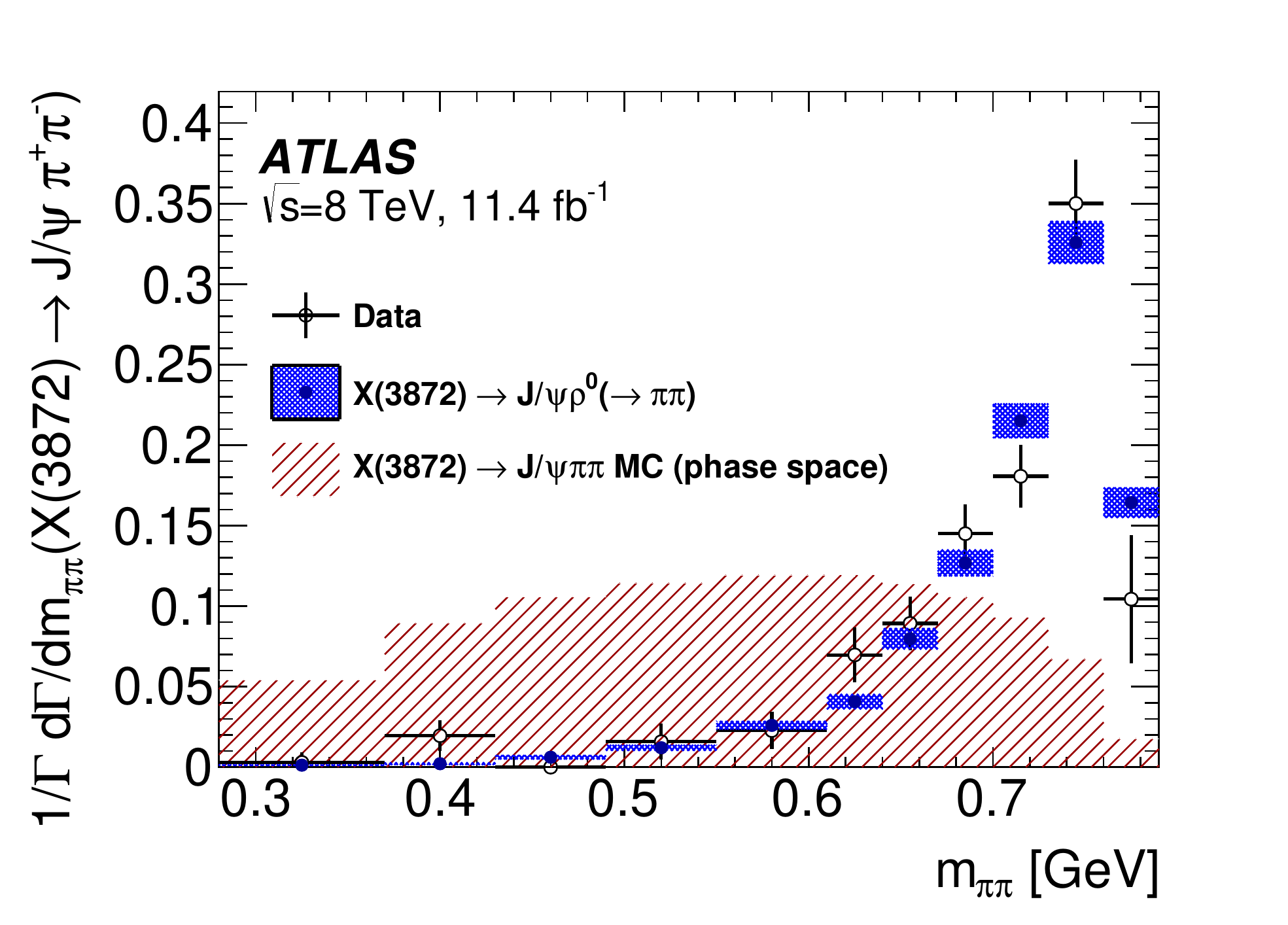}
\caption{(Left) Normalised differential decay width of $\psi(2S) \to J/\psi(\to \mu^+\mu^-)\pi^+\pi^-$ in bins of dipion invariant mass over the range $0.280 < m_{\pi\pi} < 0.595\,$GeV, fitted with the Voloshin--Zakharov model~\cite{atlx3872}. Also shown is the normalised $m_{\pi\pi}$ phase-space distribution (red shaded histogram). (Right) Normalised differential decay width of $X(3872) \to J/\psi(\to \mu^+\mu^-)\pi^+\pi^-$ in bins of dipion invariant mass over the range $0.28 < m_{\pi\pi} < 0.79\,$GeV~\cite{atlx3872}. Also shown is the MC prediction for the decay $X(3872) \to J/\psi(\to \mu^+\mu^-)\rho^0(\to \pi^+\pi^-)$ (blue histogram) and the normalised distribution of $m_{\pi\pi}$ phase-space (red shaded histogram).
}
\label{fig:fig_x3872_mpipi}
\end{figure}

The distributions of the dipion invariant mass
$m_{\pi\pi}$
in the
$\psi(2S)\to J/\psi\pi^+\pi^-$ and
$X(3872)\to J/\psi\pi^+\pi^-$
decays are measured by determining the corrected yields of
$\psi(2S)$ and
$X(3872)$ signals in narrow bins of
$m_{\pi\pi}$.
Figure~\ref{fig:fig_x3872_mpipi} shows
normalised differential decay widths of
$\psi(2S) \to J/\psi(\to \mu^+\mu^-)\pi^+\pi^-$ and
$X(3872) \to J/\psi(\to \mu^+\mu^-)\pi^+\pi^-$ in bins of dipion invariant mass.
The distribution for $\psi(2S)$
is well described by the Voloshin--Zakharov model~\cite{volzah}.
For $X(3872)$,
the normalised differential decay width
in bins of $m_{\pi\pi}$
is well described by MC simulation
of the $X(3872) \to J/\psi(\to \mu^+\mu^-)\rho^0(\to \pi^+\pi^-)$ decay.

\section{Search  for  a  hidden-beauty  analogue  of  the
$X(3872)$, $X_b$}

\begin{figure}[htb]
\centering
\includegraphics[height=2in]{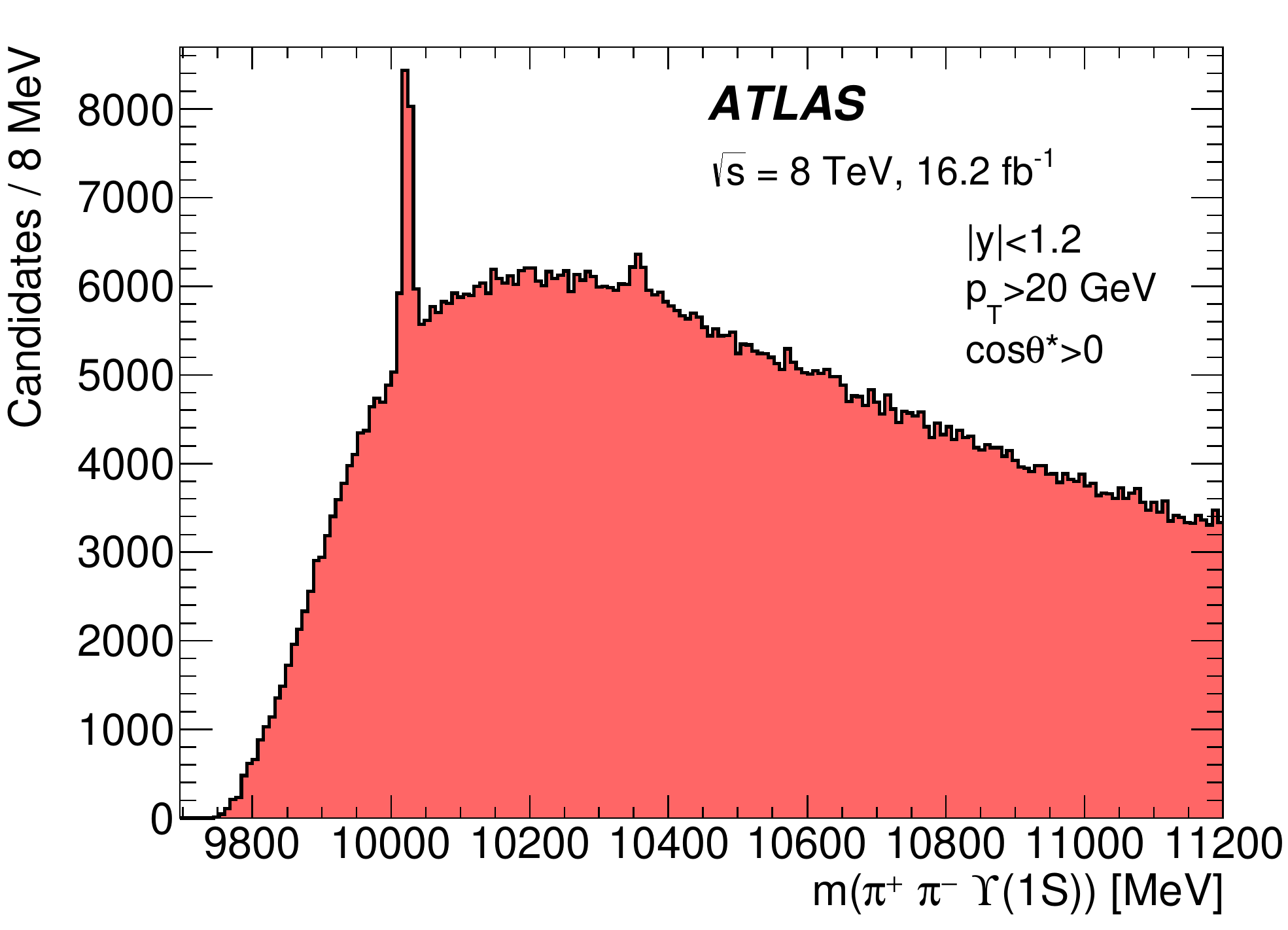}
\includegraphics[height=2in]{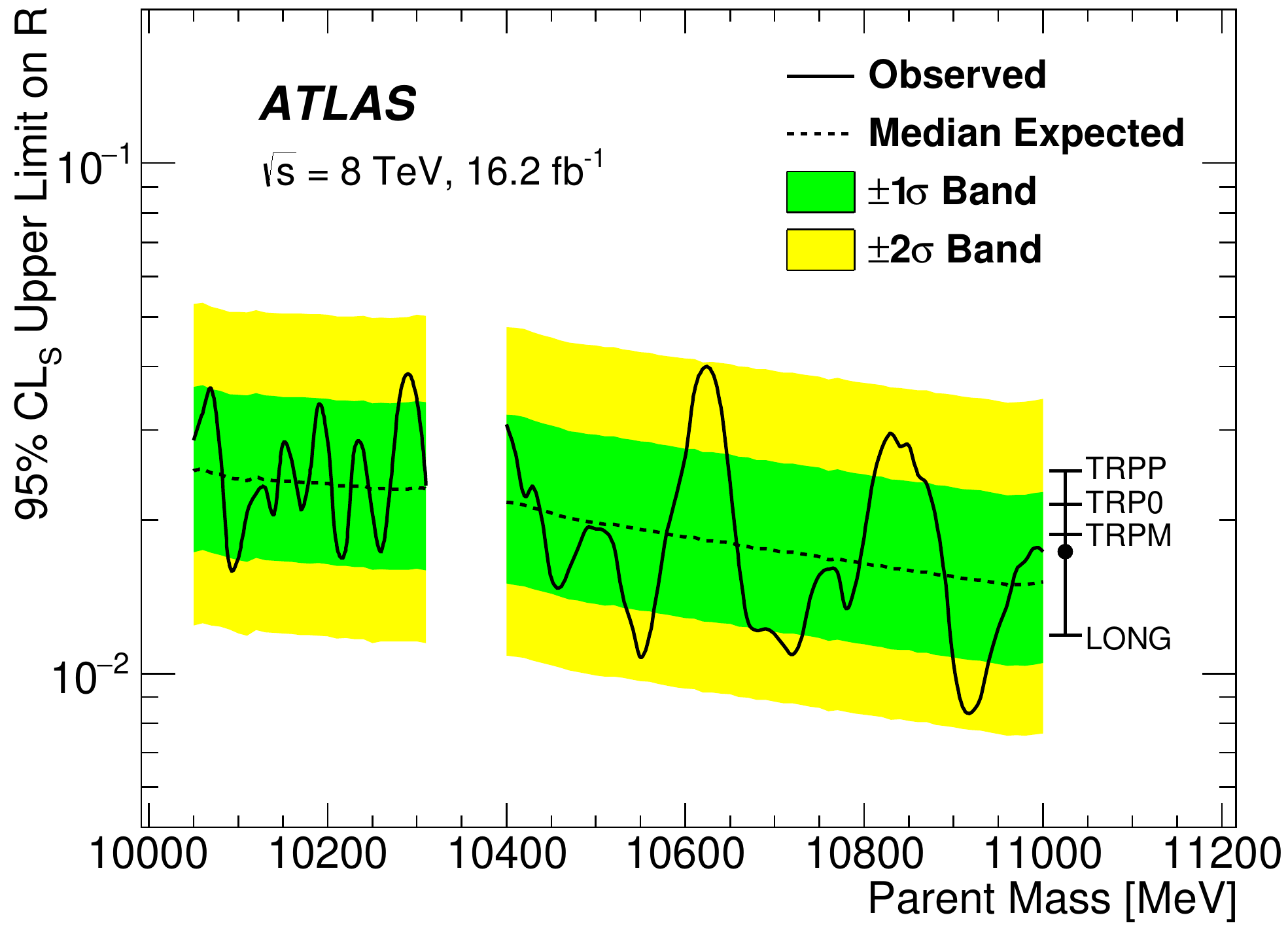}
\caption{(Left) The $\pi^+\pi^- \Upsilon(1S)$ invariant mass distribution in the kinematic bin most sensitive to an $X_b$ signal: $|y| < 1.2$, $\pt > 20\,$GeV, and $\cos\theta^* > 0$~\cite{atlxb}. 
  (Right) Observed $95\%$ CL$_S$ upper limits (solid line) on the relative production rate $R = (\sigma B)/(\sigma B)_{2S}$ of a hypothetical $X_b$ parent state decaying isotropically to $\pi^+\pi^- \Upsilon(1S)$, as a function of mass~\cite{atlxb}. The median expectation (dashed) and the corresponding ±1$\sigma$ and ±2$\sigma$ bands (green and yellow respectively) are also shown. The bar on the right shows typical shifts under alternative $X_b$ spin-alignment scenarios, relative to the isotropic case shown with the solid point.
}
\label{fig:fig_xb}
\end{figure}

Each pair of  oppositely charged muons is subjected to a common vertex fit.
Any dimuon with
an invariant mass within $350\,$MeV of the
$\Upsilon(1S)$ mass
is  retained  and  considered  an
$\Upsilon(1S) \to \mu^+\mu^-$
candidate. 
Dipion candidates are formed
from oppositely charged pions.
The $\Upsilon(1S)$
candidate and the dipion system are combined
by performing a four-track common-vertex fit, with the
$\mu^+\mu^-$
mass  constrained  to
the $\Upsilon(1S)$ mass.

Figure~\ref{fig:fig_xb}(Left) shows
the $\pi^+\pi^- \Upsilon(1S)$ invariant mass distribution in the kinematic bin most sensitive to an $X_b$ signal: $|y| < 1.2$, $\pt > 20\,$GeV, and $\cos\theta^* > 0$.
The angle $\theta^*$ is
defined as the angle, in the parent rest frame, between the dipion momentum and
the lab-frame parent momentum.
The only apparent peaks are at the masses of the $\Upsilon(2S)$ (10023$\,$MeV) and $\Upsilon(3S)$ (10355$\,$MeV).
A  hypothesis  test  for  the  presence  of  a narrow
$X_b$
peak  is  performed every $10\,$MeV from $10\,$GeV to $11\,$GeV.
At each mass, the presence of a signal is tested
by performing simultaneous fits to the nearby
$\pi^+\pi^- \Upsilon(1S)$ mass
spectrum in these bins;  no evidence for new narrow states is
found for masses $10.05 - 10.31\,$GeV and $10.40 - 11.00\,$GeV.
Upper limits are also set on the ratio
$R=
[\sigma (pp\to X_b)\bran(X_b\to\pi^+\pi^- \Upsilon(1S))]/
[\sigma (pp\to \Upsilon(2S))\bran(\Upsilon(2S)\to\pi^+\pi^- \Upsilon(1S))]$.
Figure~\ref{fig:fig_xb}(Right) shows
the observed $95\%$ CL$_S$ upper limits on the relative production rate.
The results range from $0.8\%$ to $4.0\%$ depending on the
$X_b$ mass.
The  analogous  ratio  for  the
$X(3872)$  is  $6.56\%$:
%a  value  this large
such or larger value
is excluded for all
$X_b$ masses considered.

\section{Summary}

The $\lb \rightarrow \psi(2S) \Lambda^0$ decay
is observed
and
the branching ratio
of the
$\lb \rightarrow \psi(2S) \Lambda^0$ and
$\lb \rightarrow J/\psi \Lambda^0$ decays is measured.
%~\cite{atllbpsi2s}.
The decays
$B_c^+\to J/\psi D_s^+$ and
$B_c^+\to J/\psi D_s^{*+}$ are studied
and their brnaching fractions are measured relative
to that of the $B_c^+\to J/\psi \pi^+$ decay.
%~\cite{atlbcds}.
The production cross sections and properties of
the hidden-charm states $X(3872)$ and $\psi(2S)$ are measured
in their decays to $J/\psi \pi^+\pi^-$.
%~\cite{atlx3872}.
A  search  for  a  hidden-beauty  analogue  of  the
$X(3872)$, $X_b$,  is
conducted  by  reconstructing
$\Upsilon(1S)(\to \mu^+\mu^-) \pi^+\pi^-$ events.
%~\cite{atlxb}.

%%%%%%%%%%%%%%%%%%%%%%%%%%%%%%%%%%%%%%%%%%%%%%%%%%%%%%%%%%%%%%%%%%%%%%%%%
%%
%%   use this format to include an .eps figure into your paper
%%
%\begin{figure}[htb]
%\centering
%\includegraphics[height=2in]{head_lhcp2017.jpg}
%\caption{ Place the caption here}
%\label{fig:figure1}
%\end{figure}
%%%%%%%%%%%%%%%%%%%%%%%%%%%%%%%%%%%%%%%%%%%%%%%%%%%%%%%%%%%%%%%%%%%%%%%%%%%

%See Figure \ref{fig:figure1} and Table \ref{tab:table1}. 

%%%%%%%%%%%%%%%%%%%%%%%%%%%%%%%%%%%%%%%%%%%%%%%%%%%%%%%%%%%%%%%%%%%%%%%%%
%%
%%   use this format to include a LaTeX table  into your paper
%%
%\begin{table}[t]
%\begin{center}
%\begin{tabular}{l|ccc}  
%Patient &  Initial level($\mu$g/cc) &  w. Magnet &  
%w. Magnet and Sound \\ \hline
% Guglielmo B.  &   0.12     &     0.10      &     0.001  \\
% Ferrando di N. &  0.15     &     0.11      &  $< 0.0005$ \\ \hline
%\end{tabular}
%\caption{ place the caption here }
%\label{tab:table1}
%\end{center}
%\end{table}
%%%%%%%%%%%%%%%%%%%%%%%%%%%%%%%%%%%%%%%%%%%%%%%%%%%%%%%%%%%%%%%%%%%%%%%%%%%

%%  if necessary
\Acknowledgements
Participation in the conference was supported by the Russian Foundation for Basic Research, grant 15-02-08133.


\begin{thebibliography}{99}

%%
%%  bibliographic items can be constructed using the LaTeX format in SPIRES:
%%    see    http://www.slac.stanford.edu/spires/hep/latex.html
%%  SPIRES will also supply the CITATION line information; please include it.
%%

\bibitem{atldet}
%  G.~Aad {\it et al.}  [ATLAS Collaboration],
  ATLAS Collaboration,
  JINST {\bf 3}, S08003 (2008).

\bibitem{atllbpsi2s}
%  G.~Aad {\it et al.}  [ATLAS Collaboration],
  ATLAS Collaboration,
  Phys.\ Lett.\ B {\bf 751}, 63 (2015)
  [arXiv:1507.08202 [hep-ex]].

\bibitem{atlbcds}
%  G.~Aad {\it et al.}  [ATLAS Collaboration],
  ATLAS Collaboration,
  Eur.\ Phys.\ J.\ C {\bf 76}, 1 (2016)
  [arXiv:1507.07099 [hep-ex]].

\bibitem{atlx3872}
%  M.~Aaboud {\it et al.}  [ATLAS Collaboration],
  ATLAS Collaboration,
  JHEP {\bf 01}, 117 (2017)
  [arXiv:1610.09303 [hep-ex]].

\bibitem{atlxb}
%  G.~Aad {\it et al.}  [ATLAS Collaboration],
  ATLAS Collaboration,
  Phys.\ Lett.\ B {\bf 740}, 199 (2015)
  [arXiv:1410.4409 [hep-ex]].

\bibitem{pdg2016}
  C.~Patrignani {\it et al.}  [Particle Data Group],
  Chin.\ Phys.\ C {\bf 40}, 100001 (2016).

\bibitem{modgauss}
  S.~Chekanov {\it et al.}  [ZEUS Collaboration],
  Eur.\ Phys.\ J.\ C {\bf 44}, 13 (2005)
  [arXiv:hep-ex/0505008].

\bibitem{gutsche}
T.~Gutsche {\it et al.},
Phys.\ Rev.\ D {\bf 88}, 114018 (2013)
  [arXiv:1309.7879 [hep-ph]].

\bibitem{gutsche2}
T.~Gutsche {\it et al.},
Phys.\ Rev.\ D {\bf 92}, 114008 (2015)
  [arXiv:1510.02266 [hep-ph]].

\bibitem{lhcbbcds}
  R.~Aaij {\it et al.}  [LHCb Collaboration],
  Phys.\ Rev.\ D {\bf 87}, 112012 (2013)
  [arXiv:1304.4530 [hep-ex]].

\bibitem{meng}
  C.~Meng, H.~Hang and K.-T.~Chao,
  [arXiv:1304.6710 [hep-ph]].

\bibitem{berezhnoy}
  A.~V.~Berezhnoy and A.~K.~Likhoded,
  [arXiv:1309.1979 [hep-ph]].

\bibitem{volzah}
  M.~B.~Voloshin and V.~I.~Zakharov,
  Phys.\ Rev.\ Lett.\ {\bf 45}, 688 (1980).

%\bibitem{Aad:2012tfa} 
%  G.~Aad {\it et al.}  [ATLAS Collaboration],
%  %``Observation of a new particle in the search for the Standard Model Higgs boson with the ATLAS detector at the LHC,''
%  Phys.\ Lett.\ B {\bf 716}, 1 (2012)
%  [arXiv:1207.7214 [hep-ex]].
%  %%CITATION = ARXIV:1207.7214;%%
%  %3009 citations counted in INSPIRE as of 22 Jul 2014
  
  
%%\cite{Chatrchyan:2012ufa}
%\bibitem{Chatrchyan:2012ufa} 
%  S.~Chatrchyan {\it et al.}  [CMS Collaboration],
%  %``Observation of a new boson at a mass of 125 GeV with the CMS experiment at% the LHC,''
%  Phys.\ Lett.\ B {\bf 716}, 30 (2012)
%  [arXiv:1207.7235 [hep-ex]].
%  %%CITATION = ARXIV:1207.7235;%%
%  %2951 citations counted in INSPIRE as of 22 Jul 2014



\end{thebibliography}
\end{document}